%% file: main.tex
\documentclass[copyright,creativecommons]{eptcs}

\providecommand{\event}{ThEdu'17 Post-Proceedings} 
\usepackage{underscore}           

\title{Natural Deduction and the Isabelle Proof Assistant}
\author{J{\o}rgen Villadsen
\email{jovi@dtu.dk}
\and
Andreas Halkj{\ae}r From \qquad\qquad Anders Schlichtkrull
\email{}
\institute{DTU Compute - Department of Applied Mathematics and Computer Science,\\[1ex]
Technical University of Denmark, Richard Petersens Plads, Building 324, DK-2800 Kongens Lyngby, Denmark}
}
\def\titlerunning{Natural Deduction and the Isabelle Proof Assistant}
\def\authorrunning{J. Villadsen, A. H. From \&\ A. Schlichtkrull}

\newcommand{\x}[1]{\textsf{#1}}

\newcommand{\nadea}{\x{NaDeA}}

\usepackage{amsmath}
\usepackage{textcomp}
\usepackage{wasysym}
\usepackage{listings}
\usepackage{proof}
\usepackage{tikz}
\usetikzlibrary{arrows,positioning,shapes}

\lstdefinelanguage{Isar}%
  {
  morekeywords={if,else,
also,
and,
assume,
begin,
corollary,
datatype,
end,
finally,
fix,
for,
fun,
have,
imports,
inductive,
lemma,
moreover,
next,
obtain,
oops,
primrec,
proof,
qed,
show,
then,
theorem,
theory,
type_synonym,
ultimately,
using,
where
},
        sensitive=true,
        morecomment=[l]{(*},
        literate=
   	        {~}{{$\,\neg\,$}}1
   	        {?}{{$\,\exists$}}1
	        {!}{{$\,\forall$}}1
	        {==>}{{$\,\Longrightarrow\,\:$}}1
	        {=>}{{$\,\Rightarrow\,\:$}}1
	        {\% }{{$\lambda$}}1
       }[keywords,comments,strings]%

\usepackage{isabelle, isabellesym}
\isabellestyle{it}

\newcommand{\DefineSnippet}[2]{%
   \expandafter\newcommand\csname snippet--#1\endcsname{%
     \begin{quote}
     \begin{isabelle}
     #2
     \end{isabelle}
     \end{quote}}}

\newcommand{\Snippet}[1]{\ifcsname snippet--#1\endcsname {\csname snippet--#1\endcsname} \else
  +++++++ERROR: Snippet ``#1'' not defined+++++++ \fi}
 
\input{snippets}

\begin{document}
\maketitle

\begin{abstract}
We describe our Natural Deduction Assistant (\nadea) and the interfaces between the Isabelle proof assistant and \nadea. In particular, we explain how \nadea, using a generated prover that has been verified in Isabelle, provides feedback to the student, and also how \nadea, for each formula proved by the student, provides a generated theorem that can be verified in Isabelle.
\end{abstract}

\section{Introduction}

A textbook on logic in computer science like Huth \&\ Ryan \cite{Huth} focuses on natural deduction.

\smallskip
\begin{quote}
Natural deduction is a common name for the class of proof systems composed of simple and self-evident inference rules based upon methods of proof and traditional ways of reasoning that have been applied since antiquity in deductive practice.
\end{quote}
\hfill-- Andrzej Indrzejczak, Internet Encyclopedia of Philosophy, \url{http://www.iep.utm.edu/nat-ded/}
\bigskip

\noindent
We have developed a website for teaching first-order logic and natural deduction to computer science bachelor students:

\begin{center}
\url{https://nadea.compute.dtu.dk/}
\end{center}

Our motivation is that our students should obtain a logical background that give them the prerequisites to build and formally verify dependent software. Building and verifying such software systems can be done in proof assistant computer programs like Isabelle \cite{Nipkow} in which natural deduction reasoning is a central concept. Our website therefore forms part of a course which, among other subjects, teaches natural deduction.
The website consists of a web application, \nadea, which implements a natural deduction proof system. The website also contains explanations of the system. Lastly, the website contains the ProofJudge component which is used for assessing exercises done by students in \nadea\ and providing them feedback.
The syntax, semantics and a sound and complete proof system has been formalized in the proof assistant Isabelle such that all definitions are very precise. The present paper describes the developments since our previous paper on the system \cite{IFCoLog}, in particular the interfaces to the Isabelle proof assistant. We therefore start out by giving some background on the Isabelle proof assistant.

\subsection{The Isabelle Proof Assistant}

Proof assistants are computer programs that assist computer scientists, mathematicians and logicians in defining concepts and proving properties about them. Furthermore, they ensure correctness of the proofs that are constructed.

Isabelle, or more precisely, Isabelle/HOL, is one such proof assistant. It implements higher-order logic (HOL) and we consider first-order logic (FOL) as a subset of HOL.
For instance, a computer scientist can define a sorting algorithm in Isabelle/HOL since Isabelle/HOL contains functionality for defining functions in a similar style to a functional programming language like Haskell or Standard ML. Likewise, Isabelle contains functionality for defining datatypes and more.
The computer scientists can use Isabelle/HOL to prove their algorithms correct. In Isabelle the preferred format of proofs uses the proof language Isar \cite{Isar} which looks like a mixture of English, logic and a programming language, and is based on natural deduction. Each step of the proof is specified by the user and steps are chained together using proof methods, which implement rewriting or automatic theorem proving. In each step the user specifies which proof method to apply and which lemmas the method should use. This is done either by hand or using the Sledgehammer tool which filters out relevant lemmas from the current context and employs external top-of-the-line provers such as SPASS, E, Vampire, CVC4 and Z3 to find a proof that is then automatically reconstructed in Isabelle using an appropriate proof method.

Functions written in the functional programming style can be exported from Isabelle to the languages Standard ML, Haskell, OCaml and Scala. Thus the computer scientist can obtain a verified sorting program. We use this feature too.

User interaction with Isabelle/HOL can be done in the Prover Integrated Development Environments (PIDEs) Isabelle/jEdit and Visual Studio Code.  This gives access to various kinds of feedback from the system such as information about the execution of proof methods, type inference, search for lemmas, lookup of definitions and much more. Isabelle contains a large library of definitions, functions, lemmas and theorems on which new developments can be built. Even more can be found in the Archive of Formal Proofs which is a collection of formal developments in Isabelle that anyone can submit to.

Another popular proof assistant is Coq, which also uses natural deduction.

\subsection{Overview of the Paper}

In Section 2 we explain \nadea\ and the natural deduction system that it implements. In Section 3 we explain how we formalize the syntax and semantics of first-order logic in Isabelle. In Section 4 we explain the formalization of the natural deduction system in Isabelle. In Section 5 we present our new developments, which form interfaces between Isabelle and \nadea. In particular, we explain how \nadea, using verified code generated by Isabelle, can check if the students are on the right track in their proofs, and we explain how \nadea\ can now export its proof to a format that can be checked by Isabelle. In Section 6 we discuss related work and in Section 7 we conclude.

\section{The Natural Deduction Assistant (\nadea)}

The syntax of first-order logic is based on $\bot$ for falsehood, $\rightarrow$ for implication, $\lor$ for disjunction and $\land$ for conjunction. We use the following abbreviations:
$$
\begin{array}{r@{~~~}c@{~~~}l}
\top & \equiv & \bot \rightarrow \bot \\[1ex]
\neg \phi & \equiv & \phi \rightarrow \bot \\[1ex]
\phi \leftrightarrow \psi & \equiv & (\phi \rightarrow \psi) \land (\psi \rightarrow \phi)
\end{array}
$$
These abbreviations are outside \nadea\ in order to simplify the proof system. Also equality ($=$) is omitted from \nadea. We use $\exists$ and $\forall$ for existential and universal quantification, respectively. 

\subsection{Natural Deduction --- A Textbook Presentation}

We now present the natural deduction rules as described in the literature, following Huth \&\ Ryan \cite{Huth}. The first 9 are rules for classical propositional logic and the last 4 are for first-order logic.
Intuitionistic logic can be obtained by omitting the rule \textit{PBC} (proof by contradiction, called ``Boole'' later) and adding the $\bot$-elimination rule (also known as the rule of explosion) \cite{Seldin}.
The rules are as follows:

\begin{trivlist}
\item
$$
\infer[\textit{PBC}]
{	
	~~
    \phi
	~~
}
{
	~~
	\boxed{
		\deduce[\raisebox{1.5ex}{\vdots}]{\bot}{\neg \phi}
	}
	~~
}
~~~~~~
\infer[\rightarrow E]
{
	~~
	\psi
	~~
}
{
	~~
	\phi
    ~~&~~
    \phi \rightarrow \psi
	~~
}
~~~~~~
\infer[\rightarrow I]
{
	~~
	\phi \rightarrow \psi
 	~~
}
{
	~~
	\boxed{
		\deduce[\raisebox{1.5ex}{\vdots}]{\psi}{\phi}
	}
	~~
}
$$
\par
$$
\infer[\vee E]
{
	~~
	\chi
 	~~
}
{
	~~
    \phi \vee \psi
    ~~&~~
	\boxed{
		\deduce[\raisebox{1.5ex}{\vdots}]{\chi}{\phi}
	}
    ~~&~~
	\boxed{
		\deduce[\raisebox{1.5ex}{\vdots}]{\chi}{\psi}
	}
	~~
}
~~~~~~
\infer[\vee I_1]
{
	~~
	\phi \vee \psi
 	~~
}
{
	~~
	\phi
	~~
}
~~~~~~
\infer[\vee I_2]
{
	~~
	\phi \vee \psi
 	~~
}
{
	~~
	\psi
	~~
}
$$
\par
$$
\infer[\land E_1]
{
	~~
	\phi
 	~~
}
{
	~~
	\phi \wedge \psi
	~~
}
~~~~~~
\infer[\land E_2]
{
	~~
	\psi
 	~~
}
{
	~~
	\phi \wedge \psi
	~~
}
~~~~~~
\infer[\land I]
{
	~~
	\phi \wedge \psi
 	~~
}
{
	~~
	\phi
    ~~&~~
    \psi
	~~
}
$$
\par
$$
\infer[\exists E]
{
	~~
	\chi
 	~~
}
{
	~~
	\exists x\,\phi
    ~~&~~
	\boxed{
		\deduce[\raisebox{1.5ex}{~~~~\vdots}]{~~~~\chi}{x_0~~~~\phi\left[x_0/x\right]}
	}
	~~~~~~~~
}
~~~~~~
\infer[\exists I]
{
	~~
	\exists x\,\phi
 	~~
}
{
	~~
	\phi\left[t/x\right]
	~~
}
$$
\par
$$
\infer[\forall E]
{
	~~
	\phi\left[t/x\right]
 	~~
}
{
	~~
	\forall x\,\phi
	~~
}
~~~~~~
\infer[\forall I]
{
	~~
	\forall x\,\phi
 	~~
}
{
	~~
	\boxed{
		\deduce[\raisebox{1.5ex}{~~~~\vdots}]{~~~\phi\left[x_0/x\right]}{x_0~~~~~~~}
	}
	~~
}
$$
\end{trivlist}

\

\noindent
Side conditions to rules for quantifiers:
\begin{trivlist}
\item $\exists E$: $x_0$ does not occur outside its box (and therefore not in $\chi$).
\item $\exists I$: $t$ must be free for $x$ in $\phi$.
\item $\forall E$: $t$ must be free for $x$ in $\phi$.
\item $\forall I$: $x_0$ is a new variable which does not occur outside its box.
\end{trivlist}
In addition there is a special copy rule \cite[p.~20]{Huth}:

\begin{quotation}
\noindent
A final rule is required in order to allow us to conclude a box with a formula which has already appeared earlier in the proof. [...]
The copy rule entitles us to copy formulas that appeared before, unless they depend on temporary assumptions whose box has already been closed.
\end{quotation}
The copy rule is not needed in our formalization due to the way it manages a list of assumptions.

\subsection{A Sample Natural Deduction Proof}

Proofs in \nadea\ consist of a number of lines and each line consists of the number of the line, the rule that was applied in this line and a list of assumptions (marked in square brackets) followed by a formula.
If a line contains a \currency~symbol instead of the name of a rule, the student can click on it to specify which rule to use.
Proofs are done using backward chaining. That is, the formula in each line is a goal which the student wants to prove. 
The student can prove it by clicking on \currency~and choosing a rule that has the goal as conclusion. This introduces a number of new lines, i.e. new subgoals -- namely the premises and side-conditions of the rule we chose. This forms a tree which we represent by indenting subgoals appropriately.
When there are no \currency~symbols left the proof is finished. The reason that proofs can be finished is that the rule \textit{Assume} has zero premises, so this is the rule that is used in all leaves of a finished proof.

Let us look at a small example where we prove $(\forall x. R(x,x)) \rightarrow (\forall x. \exists y. R(x,y))$. First we need to choose a rule.

\medskip

\begin{center}
\includegraphics[scale=0.3, clip = true, trim = 0 0 0 90]{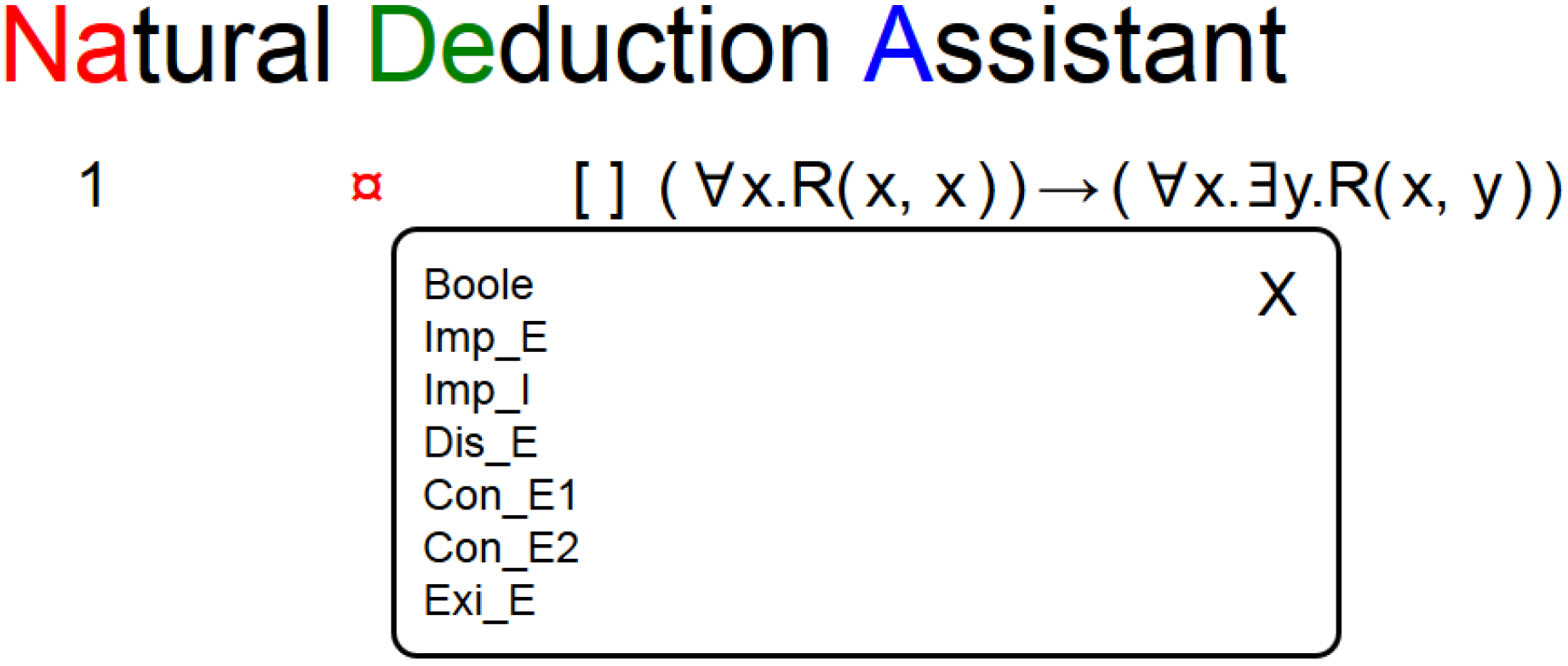}
\end{center}

Let us say that we choose \textit{Imp\_I} corresponding to Huth \&\ Ryan's $\rightarrow I$. This introduces a subgoal where, from the assumption $\forall x. R(x,x)$, we have to prove $\forall x. \exists y. R(x,y)$. Now we need to choose the rule to prove this subgoal with. 

\medskip

\begin{center}
\includegraphics[scale=0.3, clip = true, trim = 0 0 0 90]{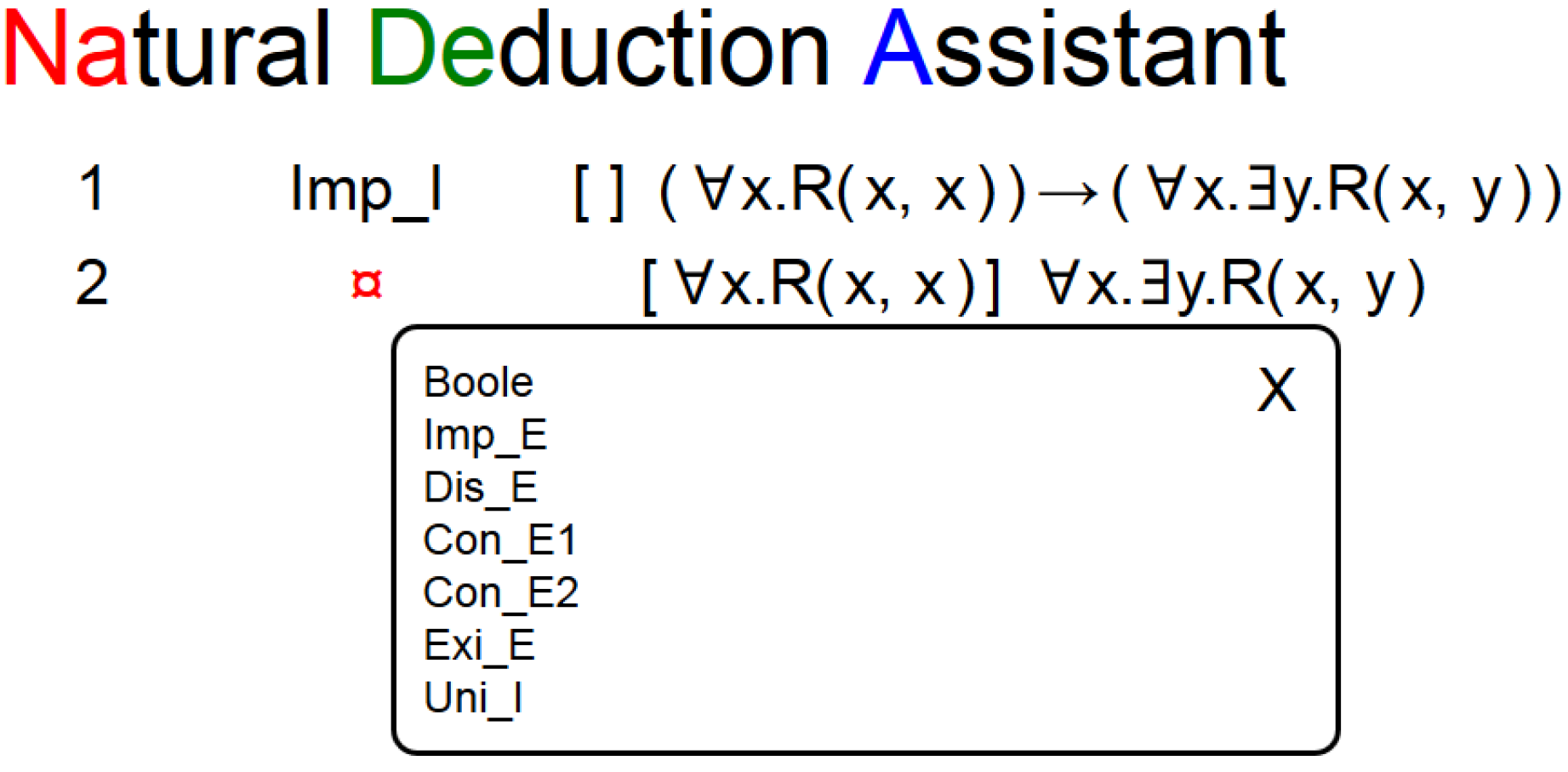}
\end{center}

We choose \textit{Uni\_I} corresponding to Huth \&\ Ryan's $\forall I$. This introduces two subgoals -- namely line 3 and line 4. Line 3 corresponds to the premise of the \textit{Uni\_I} rule -- namely that we must prove from assumption $\forall x. R(x,x)$ that $\exists x. R(c',x)$. Line 4 corresponds to the side-condition which states that $c'$ is new in the proof. The student can expand this line to see exactly what the side condition states. \nadea\ automatically chooses $c'$ such that it is new. 

We now need to choose which rule to use to prove line 3.
Since the outermost quantifier is existential, we choose the existential introduction rule \textit{Exi\_I}.
Now we need to pick the witness and \nadea\ presents us with a box for specifying a term.
We choose the constant $c'$ which then appears instead of the quantified variable in the new subgoal.

\medskip

\begin{center}
\includegraphics[scale=0.3, clip = true, trim = 0 0 0 90]{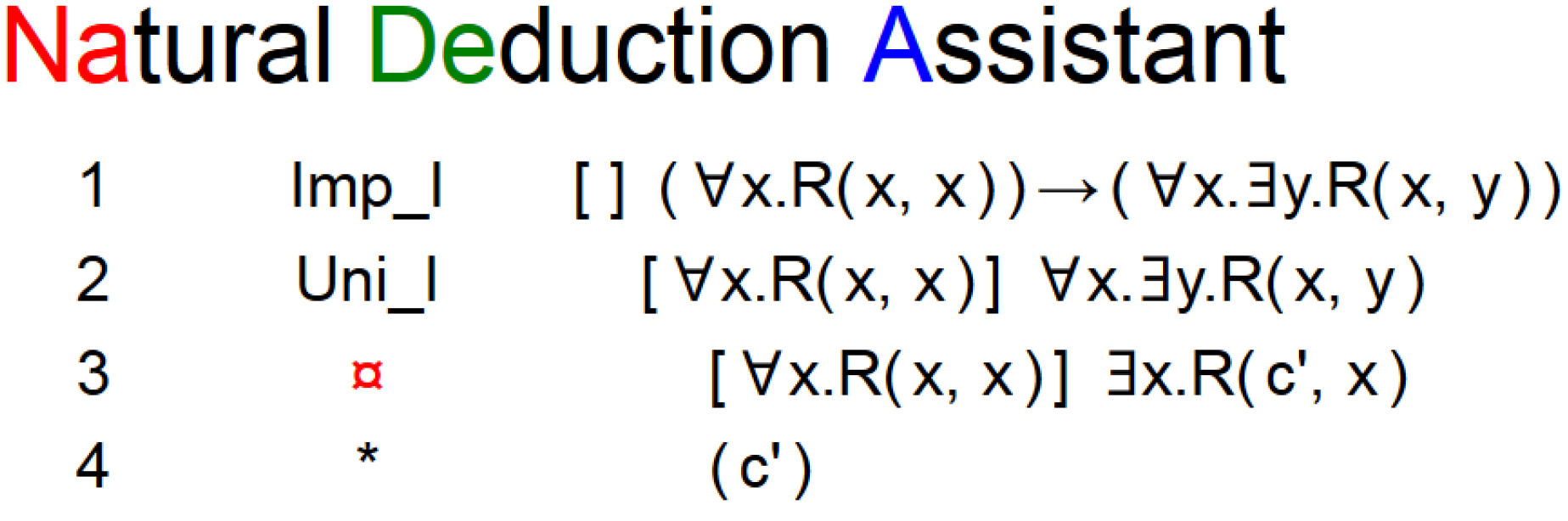}
\end{center}

Next, we choose \textit{Uni\_E} corresponding to Huth \&\ Ryan's $\forall E$. This leaves the subgoal where we from $\forall x.R(x,x)$ need to prove $\forall x.R(x,x)$. \nadea\ immediately recognizes that this can be done using the \textit{Assume} rule since the formula we want to prove appears exactly as one of the assumptions we have. This finishes the proof since there are no \currency-symbols left.

\medskip

\begin{center}
\includegraphics[scale=0.3, clip = true, trim = 0 0 0 90]{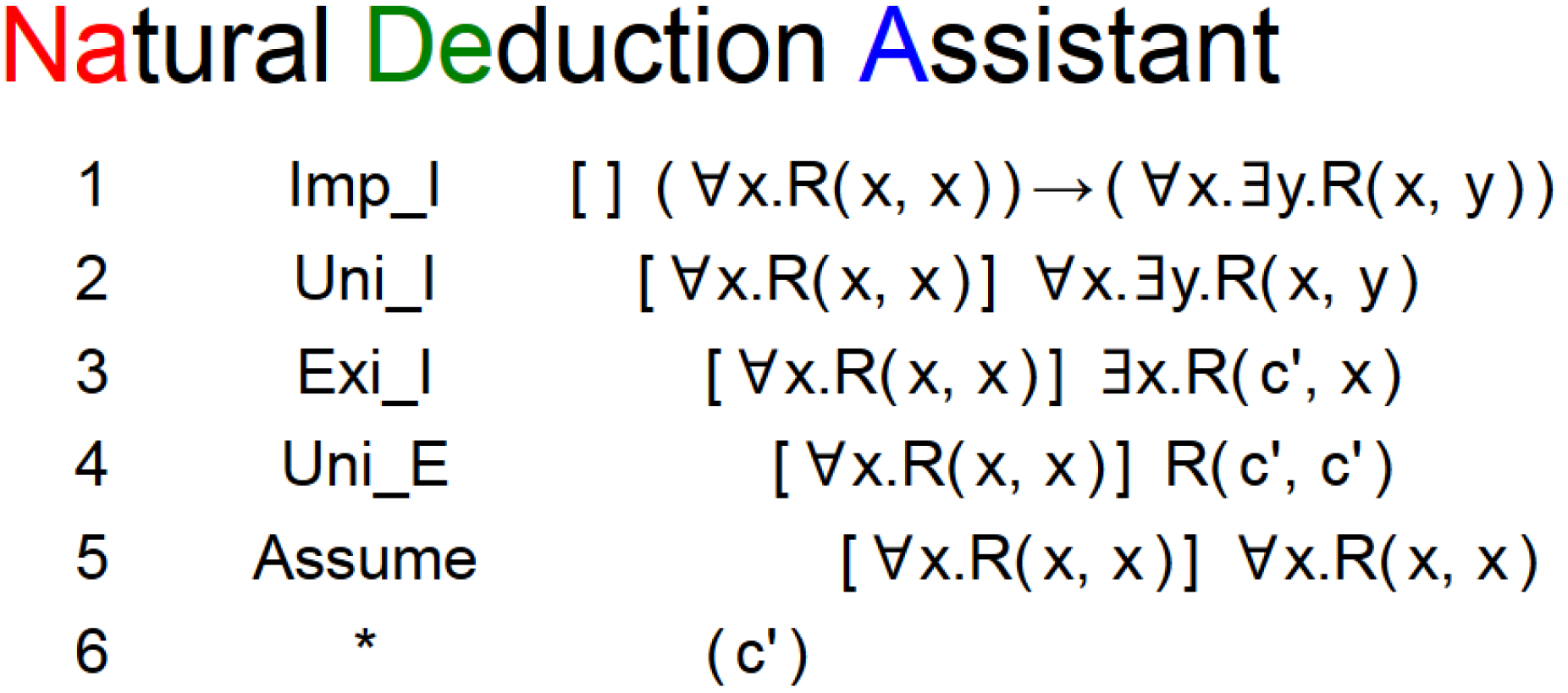}
\end{center}

\

The formula can be proved in Isabelle in a similar way but can also be proved automatically.

\section{Syntax and Semantics of First-Order Logic in Isabelle}

There are two main approaches to formalizing logics, namely shallow embeddings and deep embeddings.

\smallskip
\begin{quote}
\textbf{Shallow Embedding}\\
The act of representing one logic or language with another by providing a syntactic translation
\end{quote}
\hfill-- Wiktionary, \url{https://en.wiktionary.org/wiki/shallow_embedding}
\bigskip

\noindent
This approach is trivial since first-order logic is a subset of Isabelle/HOL.

\smallskip
\begin{quote}
\textbf{Deep Embedding}\\
The act of representing one language, typically a logic or programming language, with another by modeling expressions in the former as data in the latter
\end{quote}
\hfill-- Wiktionary, \url{https://en.wiktionary.org/wiki/deep_embedding}
\bigskip

\noindent
This approach is more involved, since we need to decide how to formalize first-order logic as data in Isabelle/HOL. The advantage, however, is that since formulas are represented as data, we can define a semantics on them and express and prove, in Isabelle, meta-theorems such as soundness and completeness of natural deduction. We therefore take this approach. Our formalization builds on a formalization by Berghofer \cite{Berghofer} according to Fitting's \cite{Fitting} approach to first-order logic.

\subsection{Syntax}

The syntax of first-order logic can be formalized using a type \textit{char list} for predicate symbols and function symbols together with datatypes for terms \textit{tm} and formulas \textit{fm}.

\begin{center}
\includegraphics[scale=0.33]{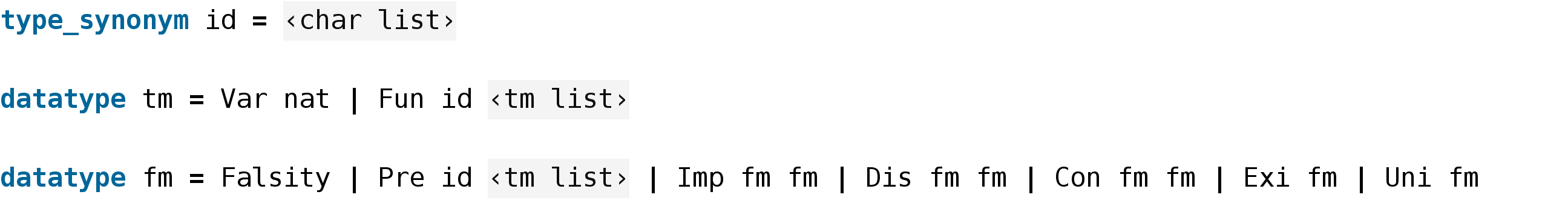}
\end{center}

Notice that variables consist of natural numbers. This is because the formalization uses de Bruijn indices to represent variables. The idea is that an occurrence of $\textit{Var} \;i$ exists in the scopes of a set of quantifiers. The occurrence is then bound to the quantifier with the $i$\,th closest scope. Here are some examples:

\[ (\forall x. \forall y. A(x, y)) \longrightarrow (\forall x. A(x, x)) \]

\begin{center}
\textit{Imp (Uni (Uni (Pre A [Var 1, Var 0]))) (Uni (Pre A [Var 0, Var 0]))}
\end{center}

\[ \forall x. \forall y. (\forall u. \forall z. A(z, u)) \longrightarrow A(x, y) \]

\begin{center}
\textit{Uni (Uni (Imp (Uni (Uni (Pre A [Var 1, Var 0]))) (Pre A [Var 1, Var 0])))}
\end{center}

\subsection{Semantics}

The semantics can be formalized as recursive functions on the syntax of first-order terms and formulas (\textit{semantics\_term} and \textit{semantics}). 
Consider the first arguments to \textit{semantics}. It represents a variable denotation and has type $\textit{nat} \Rightarrow \textit{$'a$}$, i.e. it takes a natural number (the variable) and returns an element of the universe. We represent the universe by a type variable $'a$.

\begin{center}
\includegraphics[scale=0.33]{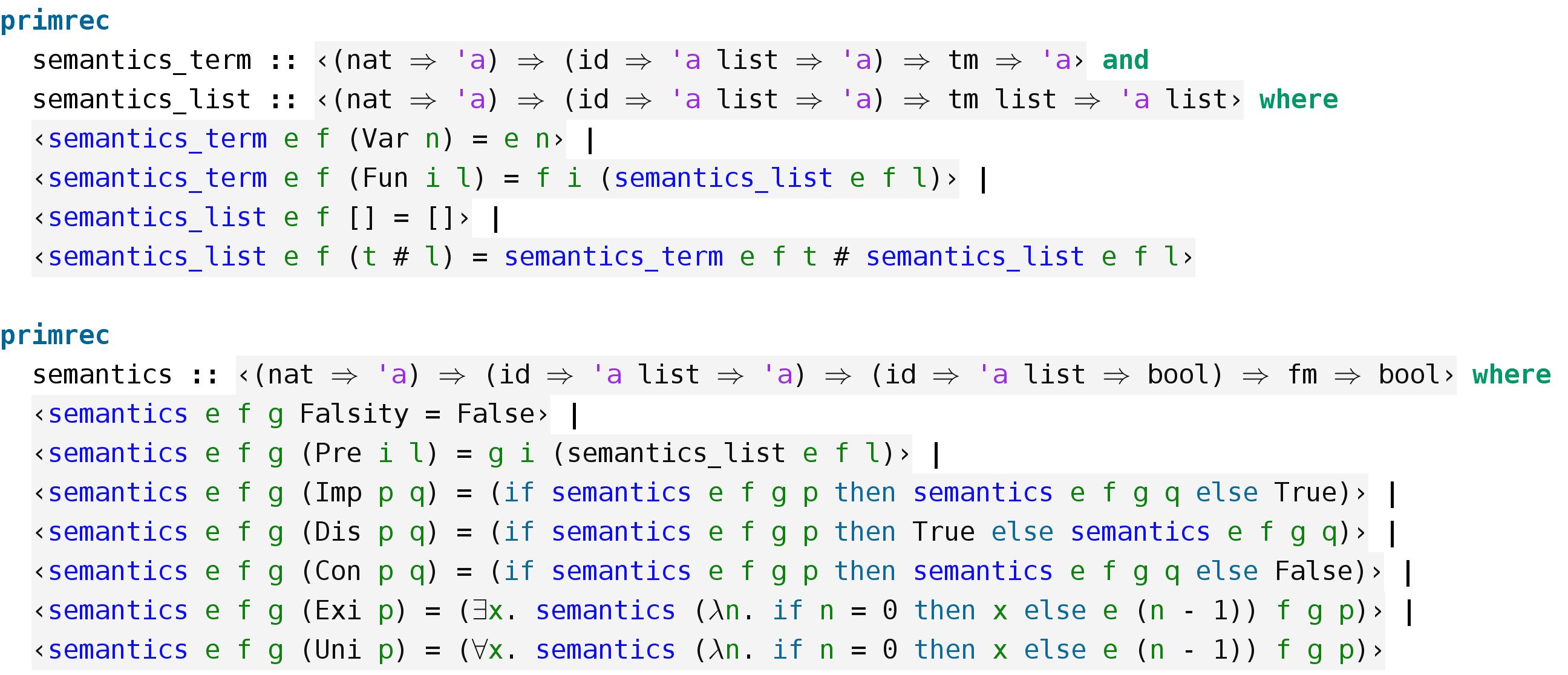}
\end{center}

\section{Natural Deduction Proof System in Isabelle}

\subsection{Membership}

Using the Isabelle command primrec we define a primitive recursive function \textit{member} that checks whether a formula (type fm) is a member of a list of formulas (type fm list).  

\begin{center}
\includegraphics[scale=0.33]{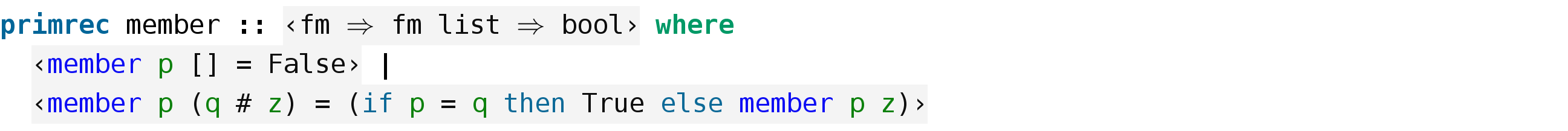}
\end{center}

This function is used in the Assume rule only.

\subsection{Newness}

The newness of a constant with respect to a term can be formalized as two mutually recursive functions which recurse on terms and lists of terms.
The functions simply check if the variable appears anywhere in the term.

Hereafter the newness of a constant with respect to a formula can be defined by recursion over formulas.

\begin{center}
\includegraphics[scale=0.33]{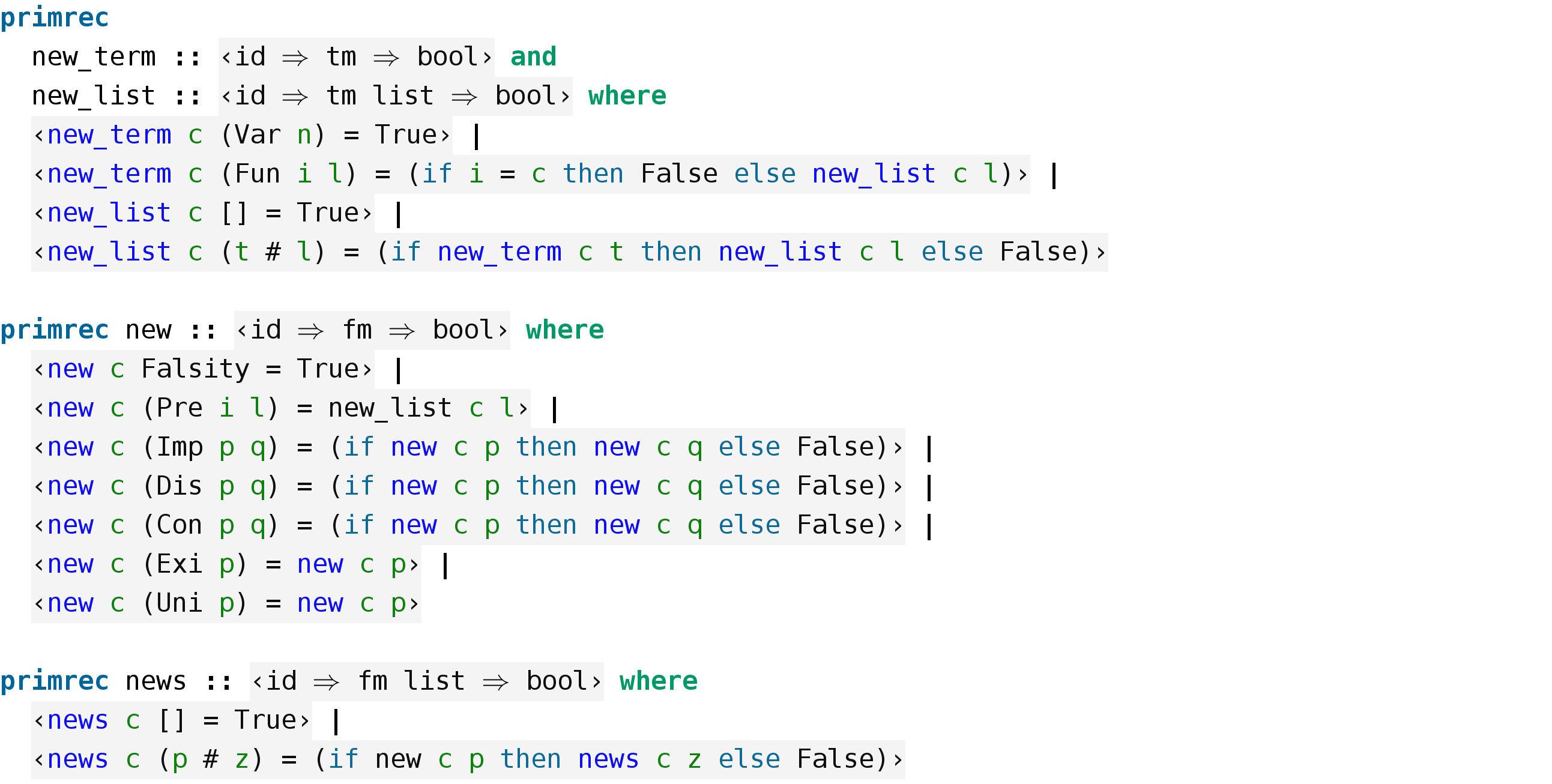}
\end{center}

The function \textit{news} checks whether a constant is new with respect to a list of formulas.

\subsection{Substitution}

Substitution is similarly defined by recursion over terms, lists of terms and formulas. In the proof system, whenever we do a substitution, we also remove a quantifier. The substitution function should take this into account. This is why the case for variables is rather involved -- it needs to make sure that the variables point to the appropriate quantifiers. For the same reason the recursion on existential and universal quantifiers add one to the variable and apply an auxiliary function \textit{inc\_term} which also adds one to all variables in the term.

\begin{center}
\includegraphics[scale=0.33]{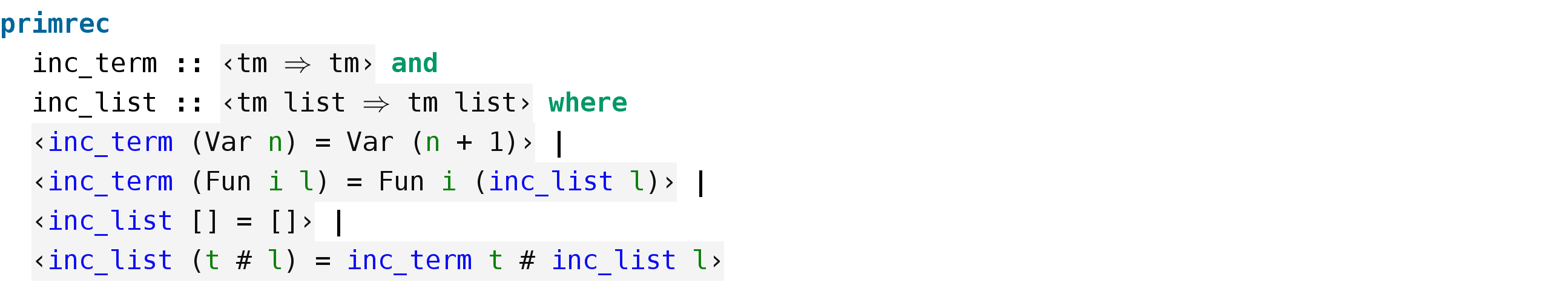}
\end{center}

\begin{center}
\includegraphics[scale=0.33]{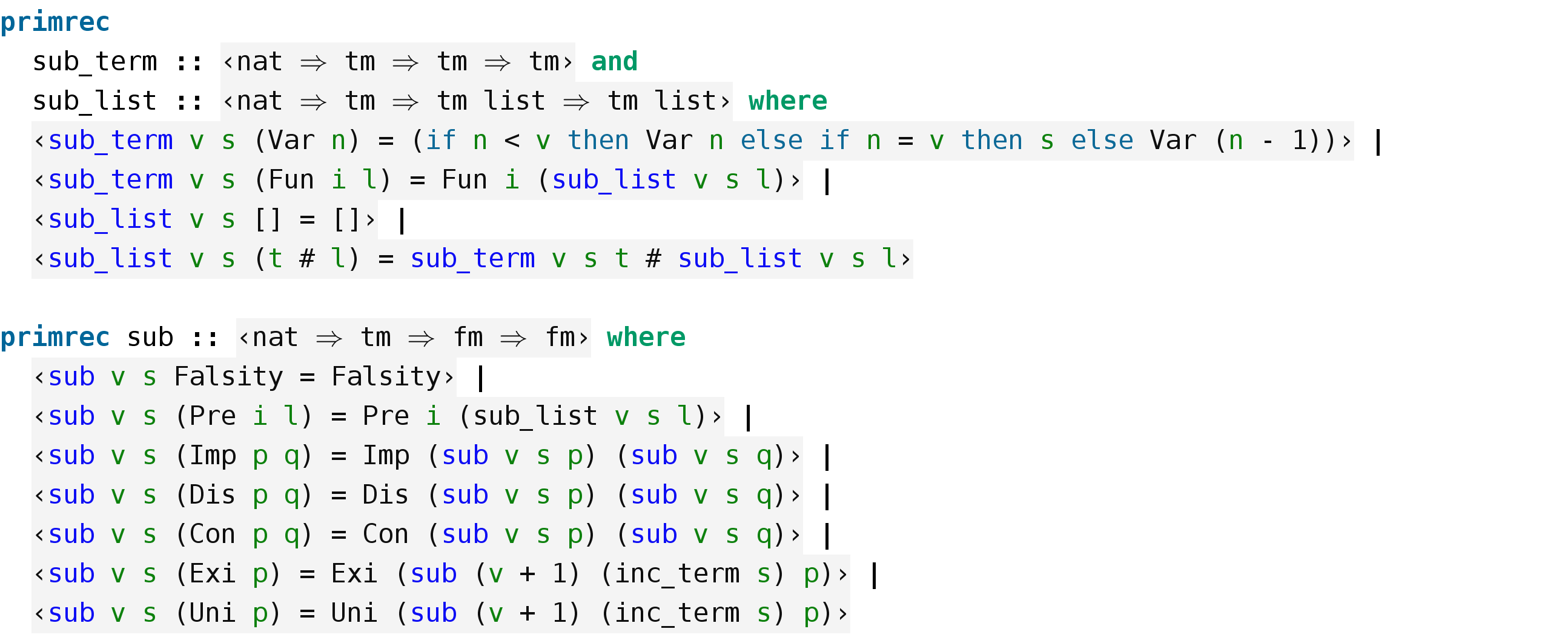}
\end{center}

\subsection{The Proof System as an Inductive Predicate}

With all these concepts in place we can define the proof system as an inductive predicate. Inductive predicates in Isabelle are specified by a number of implications that define for which elements the predicate holds. Furthermore, the predicate only holds for these elements and no others.

Each implication in our predicate corresponds to one of the rules by Huth \&\ Ryan. However, compared to those rules we keep track of assumptions by explicitly representing them in lists.

\begin{center}
\includegraphics[scale=0.33]{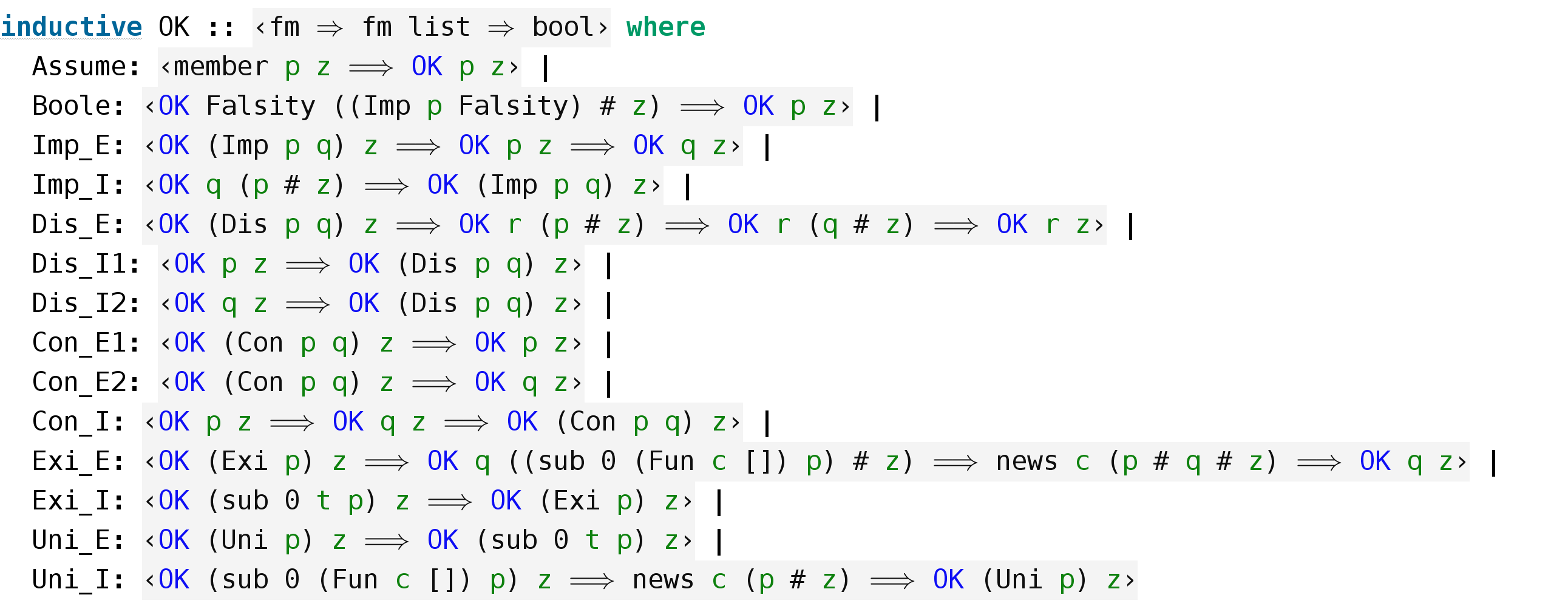}
\end{center}

\subsection{Soundness and Completeness}

We define the validity of a formula as the formula evaluating to \textit{True} in all variable denotations ($e$), function denotations ($f$) and predicate denotations ($g$) with the natural numbers as universe. This is different from the usual notion of validity which considers all universes -- not only that of the natural numbers. We can, however, prove in Isabelle that our notion of validity implies the truth of a formula in any variable denotation, function denotation and predicate denotation -- and thus the formula must indeed be valid with respect to the usual notion of validity.

We prove that the valid formulas are exactly the same as those which can be proved using \textit{OK}, hence the natural deduction proof system is sound and complete.

\begin{center}
\includegraphics[scale=0.33]{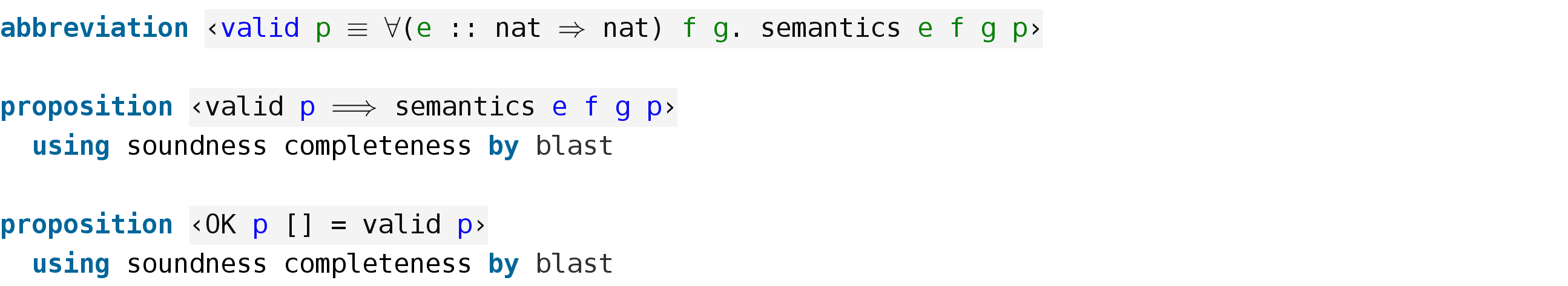}
\end{center}

\textit{OK} is preferred to longer words like \textit{Provable} or symbols like $\vdash$ mainly because it is easier to pronounce in class.
The complete Isabelle theory file with proofs of \textit{soundness} and \textit{completeness} is checked in a few seconds. The more than 4000 lines are available on the \nadea\ website.

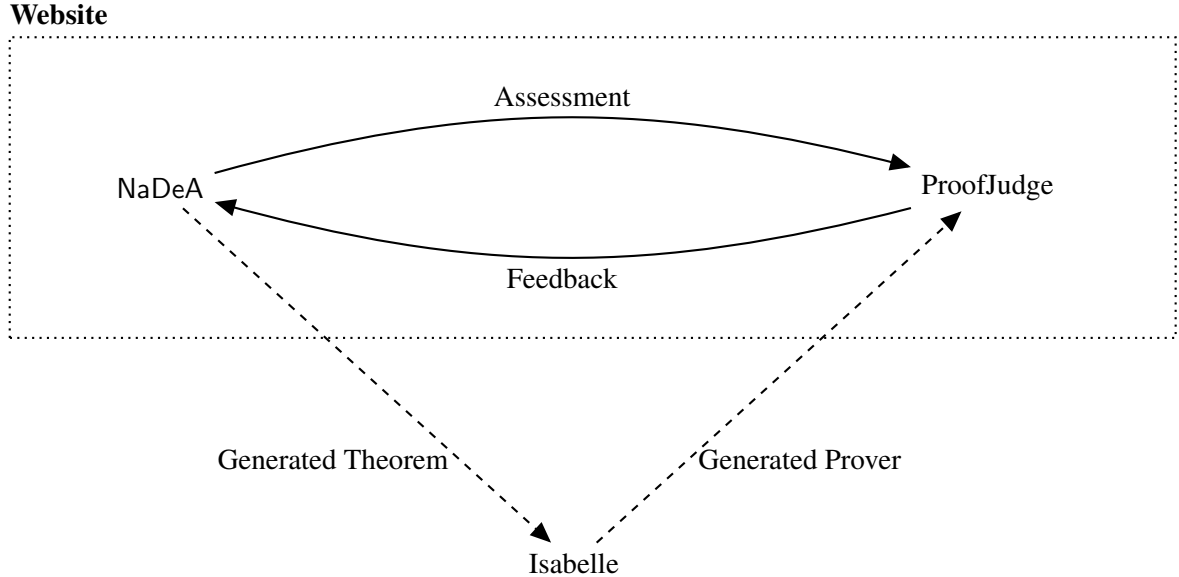
\begin{figure*}

\centering

\tikzset{%
  block/.style    = {draw, thick, rectangle, minimum height = 3em, minimum width = 13em},
  cc/.style      = {draw, thick, rectangle, rounded corners=1em, minimum height = 2em}, 
  cir/.style	= {draw, thick, circle},
  el/.style      = {draw, ellipse, minimum height = 3em}, 
  input/.style    = {coordinate},
  output/.style   = {coordinate}
}

\begin{tikzpicture}[auto, thick, node distance=2cm, >=triangle 45]
\draw [thick, dotted] (0,5) rectangle (15.5,9);
\node at (0.65,9.3) {\textbf{Website}};

\draw
	node at ( 2 , 7) [name=nadea]      {\nadea}
    node at (13 , 7) [name=proofjudge] {ProofJudge}
    node at (7.5, 2) [name=isabelle]   {Isabelle}
;

\path[->]
  (nadea)      edge [bend left = 15] node {Assessment} (proofjudge)
  (proofjudge) edge [bend left = 15] node {Feedback}   (nadea)
  (nadea)      edge [left, dashed]   node [near end]   {Generated Theorem} (isabelle)
  (isabelle)   edge [right, dashed]  node [near start] {Generated Prover}  (proofjudge)
;

\clip (0,0) rectangle (15.5, 9);

\end{tikzpicture}
\vspace*{-5ex}
\caption{\nadea, ProofJudge and Isabelle}
\vspace*{3ex}
\label{development}
\end{figure*}

\section{Verification}
We now describe our two new additions to our website which connect it to Isabelle.
This relationship is depicted in Figure \ref{development}.
Our website consists of two subcomponents: \nadea\ and ProofJudge. \nadea\ is the application in which students do their natural deduction proofs, and ProofJudge is a system that gives teachers and teaching assistants the possibility to assess the proofs done by students. The ProofJudge system also has features for providing automatic feedback.

In this section we will explain first the \emph{generated prover} relationship between ProofJudge and Isabelle and thereafter the \emph{generated theorem} relationship.

\subsection{Generated Prover}
ProofJudge can provide automatic live feedback to students by indicating whether they are on the right track in their proof attempt. It works by running a first-order prover on each subgoal to see if it can prove the subgoal. If it cannot, then this is a warning sign to the student that they may be going down a wrong path. This warning sign is indicated to the student by coloring the number of the subgoal in orange.

The prover we run is a tableau prover which is based on a logical kernel whose soundness we verified in Isabelle \cite{FOLHarrisonAFP}. 
The kernel reimplements the kernel in Harrison's textbook \cite{Harrison} and the tableau prover is a port from OCaml to SML of the tableau prover in his book.
More precisely, we define an axiomatic proof system in Isabelle/HOL and prove it sound. Hereafter Isabelle/HOL generates an SML program which forms a logical kernel. The tableau prover can be considered sound since all its logical operations are required to go through the kernel, and thus soundness is inherited from the kernel. Harrison, informally, argues that the tableau prover is also complete.
In order to get the prover to run in the browser, we use a compiler to translate it from SML to JavaScript \cite{Elsman}.

\subsection{Generated Theorem}
Formal verification of JavaScript applications is still at an infant stage -- see e.g. the work by Park, Stef\u{a}nescu and Ro\c{s}u \cite{kjs} for a promising first step. Therefore we do not attempt to verify 
\nadea\ directly.
Hence it is possible, though unlikely, that some proof rule in some context is applicable in \nadea\ when, according to the formal proof system, it should not be. 
In turn this makes it possible that an invalid formula could be proved in \nadea.

The formalization of \nadea\ in Isabelle does not suffer from this problem.
Every application of a rule is checked by Isabelle, ensuring that every side condition etc. is satisfied or subsequently discharged, assuming we trust Isabelle.
Furthermore this formalization has been proved sound, meaning that if a formula can be derived then it is valid.
We want to take advantage of this formalization to allow students to check that a formula they have proved in \nadea\ is actually valid according to the semantics, as explained above.

\subsubsection{Soundness}

When a proof has been completed in \nadea, the count of subgoals left turns into a smiley instead of a zero.
Clicking this smiley reveals a window with several tabs and an explanation of each of them.
The most basic tab is labeled \emph{Verify natural deduction proof in Isabelle} and is automatically filled with two things: An Isabelle \isacommand{proposition} with the proved formula in Isabelle syntax (shallow embedding) and a \isacommand{theorem} stating the validity of the formula using the \nadea\ syntax (deep embedding).

Consider the formula:
\[ (\forall x. \forall y. A(x, y)) \longrightarrow (\forall x. A(x, x)) \]

The generated \isacommand{proposition} for this formula is:

\Snippet{ex1prop}

We use ASCII characters for the connectives instead of Isabelle's full syntax: An exclamation mark for the universal quantifier and a three-piece arrow for the implication.
This ``typewriter'' philosophy is to make it look more like programming and thus more familiar to the students.

Of more interest is the translation of the proof itself to the Isabelle-encoded proof rules.
The \emph{soundness} theorem is used to prove the validity of the formula given its (translated) derivation:

\Snippet{ex1semantics}

The proof strongly resembles the one made in \nadea\ visually and structurally with a few complications explained shortly.
Importantly the application of rules is in the same order as on the website, allowing the student to recognize their original proof in its reformulation.

This generated proof can be copied into the end of \nadea.thy for Isabelle to verify, \nadea.thy being the Isabelle formalization of \nadea.
The student can inspect the proof itself to be convinced that it verifies the correct thing, which can also serve as an introduction to the Isabelle syntax.
We discuss some of the complications we faced when writing the program that generates the translated derivations.

\paragraph{Complications}
The propositional part of the proof, where the rules have no side conditions, can be translated directly to applications of the corresponding rules encoded in Isabelle.
For the four quantifier rules however, the substitutions and side conditions add some extra complexity.

The complications with substitutions can be seen in the proof above, in the lines proved with \isacommand{by} \textit{simp}.
These lines rewrite the state of the proof slightly and are necessary when a proof rule either assumes or produces a formula with a substitution.
Then for the (next) proof rule to apply, we need to ask Isabelle explicitly to (un)do  the substitution using the simplifier.
This adds a bit of noise to the proof but is a very simple transformation, that can be easily understood.

The second complication is, that while the website automatically checks and discharges the side conditions of proof rules, i.e. newness, this must be done explicitly when encoding them in Isabelle.
Luckily these side conditions are simple and can be discharged unobtrusively with the \isacommand{qed} \textit{simp}-syntax that runs the simplifier on the remaining goals. 

\subsubsection{Completeness}

The \nadea\ proof rules as formalized in Isabelle have been proved complete for both closed~\cite{From} and open formulas.
Nevertheless students might be interested, having derived a closed formula, in verifying the validity of its corresponding ``opened'' versions, for instance to better understand the role of environments in the semantics.
This functionality is available in the tab \emph{Open formulas - Scratch.thy} whose content is explained below.

Consider the formula:
\[ \forall x. \forall y. (\forall u. \forall z. A(z, u)) \longrightarrow A(x, y) \]
This formula is valid and thus derivable in \nadea.
It is the universal closure of:
\[ (\forall u. \forall z. A(z, u)) \longrightarrow A(x, y) \]
The student might also be interested in the validity of this formula without having to do the proof over again.
The goal is then to provide a proof of validity for the latter formula, given the derivation of the former.

We start off with a theory declaration importing the base theory:

\Snippet{ex2theory}

Again we show the formula in Isabelle syntax (shallow embedding).
This time we use the full Isabelle syntax, not just the ``typewriter'' subset, as the problem of open formulas is considered more advanced:

\Snippet{ex2prop}

Again the derivation is translated to Isabelle (deep embedding), this time as its own \isacommand{lemma} since we will need it more than once.
The derivation itself is omitted for brevity and is similar to the one above:

\Snippet{ex2given}

We show that the formula is valid using this derivation and the soundness theorem as before:

\Snippet{ex2semantics}

Our function \textit{put-unis m} used below takes a formula and puts $m$ universal quantifiers in front of it.
Note that its argument here is the open version of the original formula:

\Snippet{ex2any}

The fact \textit{any-unis} is explained below.

Given this lemma we can take $m=2$ for the validity of the original formula and $m=1$ or $m=0$ for the two possible open versions. 
The latter is what the two corollaries below do:

\Snippet{ex2cor1}

\Snippet{ex2cor2}

One such corollary is produced for every outer universal quantifier in the given formula.

Finally the theory is completed:

\Snippet{ex2end}

\paragraph{Theorem \textit{any-unis}}

This theorem is a result of having completeness for open formulas and states that given a proof with some number $k$ of outer universal quantifiers, a proof with $m$ quantifiers, either fewer or more, can be derived.

\Snippet{anyunis}

The theorem follows from our two results \textit{remove-unis}, that we can always derive a formula without its quantifiers, and \textit{put-unis}, that we can put any number back in front.
Their proofs are omitted for brevity.

\subsubsection{The Theorem-Generating Program}

Having seen examples of the theorems generated by our program we turn briefly to its implementation.
The program is written in Standard ML and, like the generated prover, translated to JavaScript for integration with \nadea.
The program is meant to be used via \nadea, but the students can also run the Standard ML code inside Isabelle's ML environment if they prefer to do so.

\nadea\ supports exporting proofs in a textual format for sharing and archival purposes.
The program parses this format into the following corresponding ML datatype:

\begin{lstlisting}[language=ml]
datatype proof = Proof of rule * fm * fm list * tm option * proof list
\end{lstlisting}

Here \textit{rule} is an enumeration type describing the rules and \textit{fm} and \textit{tm} are formulas and terms respectively.
The single formula is the result of applying the rule, called the \emph{goal}, while the list of formulas is the current assumptions.
The optional term is the constant used for the \textit{Exi\_E} and \textit{Uni\_I} rules.
Finally the list of (smaller) proofs corresponds to the premises of the rule.

This datatype is then translated to Isar by recursion on the list of premises until the \textit{Assume} case is reached.
As explained above, the applications of propositional rules are easily translated, as the way \nadea\ handles these rules corresponds exactly to the Isabelle formalization.

In the \textit{Exi\_I} and \textit{Uni\_E} cases, we need to employ a minimal version of unification that determines what constant was used in the substitution.
This amounts to traversing the current goal and the goal of the (single) premise in parallel to see what \textit{Var 0} was substituted for.
Note that after crossing a quantifier we need to consider variable 1 instead and so forth, symmetrically to how substitution works.

Knowing this allows us to (un)do the substitution before applying the rule, like in the previous example, to make the (premise) goal match the form of the premise or goal in the inductive definition.

Formally verifying the correctness of this program, as we have done for the generated prover, would require formalizing the input, the \nadea\ format, as well as the output, a fragment of Isar.
We note that even if this program wrongly translates some proofs --- or they are not really proofs --- Isabelle will in the worst case refuse to verify them.

\section{Related Work}

Our formalization builds on a formalization by Berghofer \cite{Berghofer} but the formalizations and developments described in the present paper are new.

There are several other assistants for doing proofs in logic such as PANDA \cite{panda}, Pandora \cite{pandora}, ProofWeb \cite{proofweb} and the Incredible Proof Machine \cite{breitner}.
Closest to our approach are ProofWeb and The Incredible Proof Machine since they also use aspects of proof assistants. ProofWeb does this by providing interaction between proof assistants (Coq, Isabelle and Lego) and a web interface. \nadea\ focuses on natural deduction and so does the Incredible Proof Machine where proofs are performed in a novel graphical representation forming directed acyclic graphs. However, there is no direct interface with a proof assistant, but Breitner and Lohner \cite{IncredibleProofMachineAFP} formalize in Isabelle/HOL the correspondence between their graphs and more classical proof trees, building on Blanchette, Popescu and Traytel's \cite{Blanchette} abstract approach to formalizing soundness and completeness.

The idea of formalizing logics goes beyond the first-order case. Kumar, Arthan, Myreen and Owens \cite{Kumar} formalize higher-order logic with definitions and prove it sound. Likewise, the idea of using proof assistants in teaching is not limited to teaching logics. Nipkow and Klein's textbook \cite{Klein} on programming language semantics teaches the topic on top of Isabelle.

\section{Conclusion}

This paper explains how we extended \nadea\ with two features that connect it to Isabelle. The first one consists of a tableau prover verified in Isabelle which tries to prove the subgoals of the student's goals and reports back whether it succeeded or not. This is supposed to indicate to the students whether they are on the right track or not. The second one exports proofs done in \nadea\ to proofs in Isabelle. The proofs are in a deep embedding of \nadea's proof system. Because of the soundness of this system we can conclude in Isabelle that the proved theorem is valid.

We have successfully classroom tested \nadea\ in a regular course with 70-80 bachelor students in computer science each year since 2015 (DTU 5 ECTS course 02156 Logical Systems and Logic Programming).
The students spend around 15 hours on lectures and exercises plus around 3 hours on a hand-in assignment.
In total more than 25 formulas are proved.
A condensed course was offered for PhD students at the European Summer School in Logic, Language and Information, University of Toulouse, France, 17-28 July 2017.

The current version of \nadea\ is 0.9.9 and after some polishing we plan to release 1.0.0 very soon.
The source code is available on GitHub (MIT open source license).
Future work includes investigation into other ways of integrating \nadea\ and Isabelle.

We hypothesize that proof assistants will be an indispensable tool for tomorrow's computer scientists and engineers. Their use requires a solid understanding of logic, and we hope that \nadea\ can serve as a safe playground where our students can gain such an understanding by studying natural deduction in a controlled environment. By integrating \nadea\ and Isabelle, we not only increase the trust in our system, but also practice what we preach -- following our hypothesis, the Isabelle proof assistant is an indispensable part of \nadea's design.

\section*{Acknowledgements}

Thanks to Stefan Berghofer, Christian Sternagel, Makarius Wenzel, Freek Wiedijk and John Bruntse Larsen for constructive comments in relation to drafts of the paper.
Alexander Birch Jensen was the lead programmer for earlier versions of \nadea\ and ProofJudge.

\bibliographystyle{eptcs}
\bibliography{generic}

\end{document}

%% file: snippets.tex
\DefineSnippet{anyunis}{
\isacommand{theorem}\isamarkupfalse%
\ any{\isacharunderscore}unis{\isacharcolon}\ {\isacartoucheopen}OK\ {\isacharparenleft}put{\isacharunderscore}unis\ k\ p{\isacharparenright}\ {\isacharbrackleft}{\isacharbrackright}\ {\isasymLongrightarrow}\ OK\ {\isacharparenleft}put{\isacharunderscore}unis\ m\ p{\isacharparenright}\ {\isacharbrackleft}{\isacharbrackright}{\isacartoucheclose}\isanewline
\isadelimproof
\ \ %
\endisadelimproof
\isatagproof
\isacommand{using}\isamarkupfalse%
\ put{\isacharunderscore}unis\ remove{\isacharunderscore}unis\ \isacommand{by}\isamarkupfalse%
\ blast%
\endisatagproof
{\isafoldproof}%
\isadelimproof
\endisadelimproof
}
\DefineSnippet{ex1prop}{
\isacommand{proposition}\isamarkupfalse%
\ {\isachardoublequoteopen}{\isacharparenleft}{\isacharbang}\ x{\isachardot}\ {\isacharparenleft}{\isacharbang}\ y{\isachardot}\ A{\isacharparenleft}x{\isacharcomma}\ y{\isacharparenright}{\isacharparenright}{\isacharparenright}\ {\isacharminus}{\isacharminus}{\isachargreater}\ {\isacharparenleft}{\isacharbang}\ x{\isachardot}\ A{\isacharparenleft}x{\isacharcomma}\ x{\isacharparenright}{\isacharparenright}{\isachardoublequoteclose}\isanewline
\isadelimproof
\ \ %
\endisadelimproof
\isatagproof
\isacommand{by}\isamarkupfalse%
\ blast%
\endisatagproof
{\isafoldproof}%
\isadelimproof
\endisadelimproof
}
\DefineSnippet{ex1semantics}{
\isacommand{theorem}\isamarkupfalse%
\ {\isachardoublequoteopen}semantics\ e\ f\ g\ {\isacharparenleft}Imp\ {\isacharparenleft}Uni\ {\isacharparenleft}Uni\ {\isacharparenleft}Pre\ {\isacharprime}{\isacharprime}A{\isacharprime}{\isacharprime}\ {\isacharbrackleft}Var\ {\isadigit{1}}{\isacharcomma}\ Var\ {\isadigit{0}}{\isacharbrackright}{\isacharparenright}{\isacharparenright}{\isacharparenright}\ {\isacharparenleft}Uni\ {\isacharparenleft}Pre\ {\isacharprime}{\isacharprime}A{\isacharprime}{\isacharprime}\ {\isacharbrackleft}Var\ {\isadigit{0}}{\isacharcomma}\ Var\ {\isadigit{0}}{\isacharbrackright}{\isacharparenright}{\isacharparenright}{\isacharparenright}{\isachardoublequoteclose}\isanewline
\isadelimproof
\endisadelimproof
\isatagproof
\isacommand{proof}\isamarkupfalse%
\ {\isacharparenleft}rule\ soundness{\isacharparenright}\isanewline
\ \ \isacommand{show}\isamarkupfalse%
\ {\isachardoublequoteopen}OK\ {\isacharparenleft}Imp\ {\isacharparenleft}Uni\ {\isacharparenleft}Uni\ {\isacharparenleft}Pre\ {\isacharprime}{\isacharprime}A{\isacharprime}{\isacharprime}\ {\isacharbrackleft}Var\ {\isadigit{1}}{\isacharcomma}\ Var\ {\isadigit{0}}{\isacharbrackright}{\isacharparenright}{\isacharparenright}{\isacharparenright}\ {\isacharparenleft}Uni\ {\isacharparenleft}Pre\ {\isacharprime}{\isacharprime}A{\isacharprime}{\isacharprime}\ {\isacharbrackleft}Var\ {\isadigit{0}}{\isacharcomma}\ Var\ {\isadigit{0}}{\isacharbrackright}{\isacharparenright}{\isacharparenright}{\isacharparenright}\ {\isacharbrackleft}{\isacharbrackright}{\isachardoublequoteclose}\isanewline
\ \ \isacommand{proof}\isamarkupfalse%
\ {\isacharparenleft}rule\ Imp{\isacharunderscore}I{\isacharparenright}\isanewline
\ \ \ \ \isacommand{show}\isamarkupfalse%
\ {\isachardoublequoteopen}OK\ {\isacharparenleft}Uni\ {\isacharparenleft}Pre\ {\isacharprime}{\isacharprime}A{\isacharprime}{\isacharprime}\ {\isacharbrackleft}Var\ {\isadigit{0}}{\isacharcomma}\ Var\ {\isadigit{0}}{\isacharbrackright}{\isacharparenright}{\isacharparenright}\ {\isacharbrackleft}Uni\ {\isacharparenleft}Uni\ {\isacharparenleft}Pre\ {\isacharprime}{\isacharprime}A{\isacharprime}{\isacharprime}\ {\isacharbrackleft}Var\ {\isadigit{1}}{\isacharcomma}\ Var\ {\isadigit{0}}{\isacharbrackright}{\isacharparenright}{\isacharparenright}{\isacharbrackright}{\isachardoublequoteclose}\isanewline
\ \ \ \ \isacommand{proof}\isamarkupfalse%
\ {\isacharparenleft}rule\ Uni{\isacharunderscore}I{\isacharparenright}\isanewline
\ \ \ \ \ \ \isacommand{have}\isamarkupfalse%
\ {\isachardoublequoteopen}OK\ {\isacharparenleft}Uni\ {\isacharparenleft}Uni\ {\isacharparenleft}Pre\ {\isacharprime}{\isacharprime}A{\isacharprime}{\isacharprime}\ {\isacharbrackleft}Var\ {\isadigit{1}}{\isacharcomma}\ Var\ {\isadigit{0}}{\isacharbrackright}{\isacharparenright}{\isacharparenright}{\isacharparenright}\ {\isacharbrackleft}Uni\ {\isacharparenleft}Uni\ {\isacharparenleft}Pre\ {\isacharprime}{\isacharprime}A{\isacharprime}{\isacharprime}\ {\isacharbrackleft}Var\ {\isadigit{1}}{\isacharcomma}\ Var\ {\isadigit{0}}{\isacharbrackright}{\isacharparenright}{\isacharparenright}{\isacharbrackright}{\isachardoublequoteclose}\isanewline
\ \ \ \ \ \ \ \ \isacommand{by}\isamarkupfalse%
\ {\isacharparenleft}rule\ Assume{\isacharparenright}\ simp\isanewline
\ \ \ \ \ \ \isacommand{then}\isamarkupfalse%
\ \isacommand{have}\isamarkupfalse%
\ {\isachardoublequoteopen}OK\ {\isacharparenleft}sub\ {\isadigit{0}}\ {\isacharparenleft}Fun\ {\isacharprime}{\isacharprime}c{\isacharasterisk}{\isacharprime}{\isacharprime}\ {\isacharbrackleft}{\isacharbrackright}{\isacharparenright}\ {\isacharparenleft}Uni\ {\isacharparenleft}Pre\ {\isacharprime}{\isacharprime}A{\isacharprime}{\isacharprime}\ {\isacharbrackleft}Var\ {\isadigit{1}}{\isacharcomma}\ Var\ {\isadigit{0}}{\isacharbrackright}{\isacharparenright}{\isacharparenright}{\isacharparenright}\ {\isacharbrackleft}Uni\ {\isacharparenleft}Uni\ {\isacharparenleft}Pre\ {\isacharprime}{\isacharprime}A{\isacharprime}{\isacharprime}\ {\isacharbrackleft}Var\ {\isadigit{1}}{\isacharcomma}\ Var\ {\isadigit{0}}{\isacharbrackright}{\isacharparenright}{\isacharparenright}{\isacharbrackright}{\isachardoublequoteclose}\isanewline
\ \ \ \ \ \ \ \ \isacommand{by}\isamarkupfalse%
\ {\isacharparenleft}rule\ Uni{\isacharunderscore}E{\isacharparenright}\isanewline
\ \ \ \ \ \ \isacommand{then}\isamarkupfalse%
\ \isacommand{have}\isamarkupfalse%
\ {\isachardoublequoteopen}OK\ {\isacharparenleft}Uni\ {\isacharparenleft}Pre\ {\isacharprime}{\isacharprime}A{\isacharprime}{\isacharprime}\ {\isacharbrackleft}Fun\ {\isacharprime}{\isacharprime}c{\isacharasterisk}{\isacharprime}{\isacharprime}\ {\isacharbrackleft}{\isacharbrackright}{\isacharcomma}\ Var\ {\isadigit{0}}{\isacharbrackright}{\isacharparenright}{\isacharparenright}\ {\isacharbrackleft}Uni\ {\isacharparenleft}Uni\ {\isacharparenleft}Pre\ {\isacharprime}{\isacharprime}A{\isacharprime}{\isacharprime}\ {\isacharbrackleft}Var\ {\isadigit{1}}{\isacharcomma}\ Var\ {\isadigit{0}}{\isacharbrackright}{\isacharparenright}{\isacharparenright}{\isacharbrackright}{\isachardoublequoteclose}\isanewline
\ \ \ \ \ \ \ \ \isacommand{by}\isamarkupfalse%
\ simp\isanewline
\ \ \ \ \ \ \isacommand{then}\isamarkupfalse%
\ \isacommand{have}\isamarkupfalse%
\ {\isachardoublequoteopen}OK\ {\isacharparenleft}sub\ {\isadigit{0}}\ {\isacharparenleft}Fun\ {\isacharprime}{\isacharprime}c{\isacharasterisk}{\isacharprime}{\isacharprime}\ {\isacharbrackleft}{\isacharbrackright}{\isacharparenright}\ {\isacharparenleft}Pre\ {\isacharprime}{\isacharprime}A{\isacharprime}{\isacharprime}\ {\isacharbrackleft}Fun\ {\isacharprime}{\isacharprime}c{\isacharasterisk}{\isacharprime}{\isacharprime}\ {\isacharbrackleft}{\isacharbrackright}{\isacharcomma}\ Var\ {\isadigit{0}}{\isacharbrackright}{\isacharparenright}{\isacharparenright}\ {\isacharbrackleft}Uni\ {\isacharparenleft}Uni\ {\isacharparenleft}Pre\ {\isacharprime}{\isacharprime}A{\isacharprime}{\isacharprime}\ {\isacharbrackleft}Var\ {\isadigit{1}}{\isacharcomma}\ Var\ {\isadigit{0}}{\isacharbrackright}{\isacharparenright}{\isacharparenright}{\isacharbrackright}{\isachardoublequoteclose}\isanewline
\ \ \ \ \ \ \ \ \isacommand{by}\isamarkupfalse%
\ {\isacharparenleft}rule\ Uni{\isacharunderscore}E{\isacharparenright}\isanewline
\ \ \ \ \ \ \isacommand{then}\isamarkupfalse%
\ \isacommand{have}\isamarkupfalse%
\ {\isachardoublequoteopen}OK\ {\isacharparenleft}Pre\ {\isacharprime}{\isacharprime}A{\isacharprime}{\isacharprime}\ {\isacharbrackleft}Fun\ {\isacharprime}{\isacharprime}c{\isacharasterisk}{\isacharprime}{\isacharprime}\ {\isacharbrackleft}{\isacharbrackright}{\isacharcomma}\ Fun\ {\isacharprime}{\isacharprime}c{\isacharasterisk}{\isacharprime}{\isacharprime}\ {\isacharbrackleft}{\isacharbrackright}{\isacharbrackright}{\isacharparenright}\ {\isacharbrackleft}Uni\ {\isacharparenleft}Uni\ {\isacharparenleft}Pre\ {\isacharprime}{\isacharprime}A{\isacharprime}{\isacharprime}\ {\isacharbrackleft}Var\ {\isadigit{1}}{\isacharcomma}\ Var\ {\isadigit{0}}{\isacharbrackright}{\isacharparenright}{\isacharparenright}{\isacharbrackright}{\isachardoublequoteclose}\isanewline
\ \ \ \ \ \ \ \ \isacommand{by}\isamarkupfalse%
\ simp\isanewline
\ \ \ \ \ \ \isacommand{then}\isamarkupfalse%
\ \isacommand{show}\isamarkupfalse%
\ {\isachardoublequoteopen}OK\ {\isacharparenleft}sub\ {\isadigit{0}}\ {\isacharparenleft}Fun\ {\isacharprime}{\isacharprime}c{\isacharasterisk}{\isacharprime}{\isacharprime}\ {\isacharbrackleft}{\isacharbrackright}{\isacharparenright}\ {\isacharparenleft}Pre\ {\isacharprime}{\isacharprime}A{\isacharprime}{\isacharprime}\ {\isacharbrackleft}Var\ {\isadigit{0}}{\isacharcomma}\ Var\ {\isadigit{0}}{\isacharbrackright}{\isacharparenright}{\isacharparenright}\ {\isacharbrackleft}Uni\ {\isacharparenleft}Uni\ {\isacharparenleft}Pre\ {\isacharprime}{\isacharprime}A{\isacharprime}{\isacharprime}\ {\isacharbrackleft}Var\ {\isadigit{1}}{\isacharcomma}\ Var\ {\isadigit{0}}{\isacharbrackright}{\isacharparenright}{\isacharparenright}{\isacharbrackright}{\isachardoublequoteclose}\isanewline
\ \ \ \ \ \ \ \ \isacommand{by}\isamarkupfalse%
\ simp\isanewline
\ \ \ \ \isacommand{qed}\isamarkupfalse%
\ simp\isanewline
\ \ \isacommand{qed}\isamarkupfalse%
\isanewline
\isacommand{qed}\isamarkupfalse%
\endisatagproof
{\isafoldproof}%
\isadelimproof
\endisadelimproof
}
\DefineSnippet{ex2theory}{
\isadelimtheory
\endisadelimtheory
\isatagtheory
\isacommand{theory}\isamarkupfalse%
\ Scratch\ \isakeyword{imports}\ Natural{\isacharunderscore}Deduction{\isacharunderscore}Assistant\ \isakeyword{begin}%
\endisatagtheory
{\isafoldtheory}%
\isadelimtheory
\endisadelimtheory
}
\DefineSnippet{ex2prop}{
\isacommand{proposition}\isamarkupfalse%
\ {\isacartoucheopen}{\isasymforall}x{\isachardot}\ {\isacharparenleft}{\isasymforall}y{\isachardot}\ {\isacharparenleft}{\isacharparenleft}{\isasymforall}z{\isachardot}\ {\isacharparenleft}{\isasymforall}u{\isachardot}\ A{\isacharparenleft}z{\isacharcomma}\ u{\isacharparenright}{\isacharparenright}{\isacharparenright}\ {\isasymlongrightarrow}\ A{\isacharparenleft}x{\isacharcomma}\ y{\isacharparenright}{\isacharparenright}{\isacharparenright}{\isacartoucheclose}\isanewline
\isadelimproof
\ \ %
\endisadelimproof
\isatagproof
\isacommand{by}\isamarkupfalse%
\ blast%
\endisatagproof
{\isafoldproof}%
\isadelimproof
\endisadelimproof
}
\DefineSnippet{ex2given}{
\isacommand{lemma}\isamarkupfalse%
\ as{\isacharunderscore}given{\isacharunderscore}by{\isacharunderscore}the{\isacharunderscore}user{\isacharcolon}\isanewline
\ \ {\isacartoucheopen}OK\ {\isacharparenleft}Uni\ {\isacharparenleft}Uni\ {\isacharparenleft}Imp\ {\isacharparenleft}Uni\ {\isacharparenleft}Uni\ {\isacharparenleft}Pre\ {\isacharprime}{\isacharprime}A{\isacharprime}{\isacharprime}\ {\isacharbrackleft}Var\ {\isadigit{1}}{\isacharcomma}\ Var\ {\isadigit{0}}{\isacharbrackright}{\isacharparenright}{\isacharparenright}{\isacharparenright}\ {\isacharparenleft}Pre\ {\isacharprime}{\isacharprime}A{\isacharprime}{\isacharprime}\ {\isacharbrackleft}Var\ {\isadigit{1}}{\isacharcomma}\ Var\ {\isadigit{0}}{\isacharbrackright}{\isacharparenright}{\isacharparenright}{\isacharparenright}{\isacharparenright}\ {\isacharbrackleft}{\isacharbrackright}{\isacartoucheclose}%
}
\DefineSnippet{ex2semantics}{
\isacommand{theorem}\isamarkupfalse%
\isanewline
\ \ {\isacartoucheopen}semantics\ e\ f\ g\ {\isacharparenleft}Uni\ {\isacharparenleft}Uni\ {\isacharparenleft}Imp\ {\isacharparenleft}Uni\ {\isacharparenleft}Uni\ {\isacharparenleft}Pre\ {\isacharprime}{\isacharprime}A{\isacharprime}{\isacharprime}\ {\isacharbrackleft}Var\ {\isadigit{1}}{\isacharcomma}\ Var\ {\isadigit{0}}{\isacharbrackright}{\isacharparenright}{\isacharparenright}{\isacharparenright}\ {\isacharparenleft}Pre\ {\isacharprime}{\isacharprime}A{\isacharprime}{\isacharprime}\ {\isacharbrackleft}Var\ {\isadigit{1}}{\isacharcomma}\ Var\ {\isadigit{0}}{\isacharbrackright}{\isacharparenright}{\isacharparenright}{\isacharparenright}{\isacharparenright}{\isacartoucheclose}\isanewline
\isadelimproof
\ \ %
\endisadelimproof
\isatagproof
\isacommand{using}\isamarkupfalse%
\ as{\isacharunderscore}given{\isacharunderscore}by{\isacharunderscore}the{\isacharunderscore}user\ soundness\ \isacommand{by}\isamarkupfalse%
\ blast%
\endisatagproof
{\isafoldproof}%
\isadelimproof
\endisadelimproof
}
\DefineSnippet{ex2any}{
\isacommand{lemma}\isamarkupfalse%
\ as{\isacharunderscore}given{\isacharunderscore}by{\isacharunderscore}the{\isacharunderscore}user{\isacharunderscore}any{\isacharunderscore}unis{\isacharcolon}\isanewline
\ \ {\isacartoucheopen}semantics\ e\ f\ g\ {\isacharparenleft}put{\isacharunderscore}unis\ m\ {\isacharparenleft}Imp\ {\isacharparenleft}Uni\ {\isacharparenleft}Uni\ {\isacharparenleft}Pre\ {\isacharprime}{\isacharprime}A{\isacharprime}{\isacharprime}\ {\isacharbrackleft}Var\ {\isadigit{1}}{\isacharcomma}\ Var\ {\isadigit{0}}{\isacharbrackright}{\isacharparenright}{\isacharparenright}{\isacharparenright}\ {\isacharparenleft}Pre\ {\isacharprime}{\isacharprime}A{\isacharprime}{\isacharprime}\ {\isacharbrackleft}Var\ {\isadigit{1}}{\isacharcomma}\ Var\ {\isadigit{0}}{\isacharbrackright}{\isacharparenright}{\isacharparenright}{\isacharparenright}{\isacartoucheclose}\isanewline
\isadelimproof
\endisadelimproof
\isatagproof
\isacommand{proof}\isamarkupfalse%
\ {\isacharminus}\isanewline
\ \ \isacommand{have}\isamarkupfalse%
\ {\isacartoucheopen}OK\ {\isacharparenleft}put{\isacharunderscore}unis\ {\isadigit{2}}\ {\isacharparenleft}Imp\ {\isacharparenleft}Uni\ {\isacharparenleft}Uni\ {\isacharparenleft}Pre\ {\isacharprime}{\isacharprime}A{\isacharprime}{\isacharprime}\ {\isacharbrackleft}Var\ {\isadigit{1}}{\isacharcomma}\ Var\ {\isadigit{0}}{\isacharbrackright}{\isacharparenright}{\isacharparenright}{\isacharparenright}\ {\isacharparenleft}Pre\ {\isacharprime}{\isacharprime}A{\isacharprime}{\isacharprime}\ {\isacharbrackleft}Var\ {\isadigit{1}}{\isacharcomma}\ Var\ {\isadigit{0}}{\isacharbrackright}{\isacharparenright}{\isacharparenright}{\isacharparenright}\ {\isacharbrackleft}{\isacharbrackright}{\isacartoucheclose}\isanewline
\ \ \ \ \isacommand{unfolding}\isamarkupfalse%
\ put{\isacharunderscore}unis{\isacharunderscore}def\ \isacommand{using}\isamarkupfalse%
\ as{\isacharunderscore}given{\isacharunderscore}by{\isacharunderscore}the{\isacharunderscore}user\ \isacommand{by}\isamarkupfalse%
\ simp\isanewline
\ \ \isacommand{then}\isamarkupfalse%
\ \isacommand{show}\isamarkupfalse%
\ {\isacharquery}thesis\isanewline
\ \ \ \ \isacommand{using}\isamarkupfalse%
\ any{\isacharunderscore}unis\ soundness\ \isacommand{by}\isamarkupfalse%
\ blast\isanewline
\isacommand{qed}\isamarkupfalse%
\endisatagproof
{\isafoldproof}%
\isadelimproof
\endisadelimproof
}
\DefineSnippet{ex2cor1}{
\isacommand{corollary}\isamarkupfalse%
\ without{\isacharunderscore}one{\isacharunderscore}Uni{\isacharcolon}\isanewline
\ \ {\isacartoucheopen}semantics\ e\ f\ g\ {\isacharparenleft}Uni\ {\isacharparenleft}Imp\ {\isacharparenleft}Uni\ {\isacharparenleft}Uni\ {\isacharparenleft}Pre\ {\isacharprime}{\isacharprime}A{\isacharprime}{\isacharprime}\ {\isacharbrackleft}Var\ {\isadigit{1}}{\isacharcomma}\ Var\ {\isadigit{0}}{\isacharbrackright}{\isacharparenright}{\isacharparenright}{\isacharparenright}\ {\isacharparenleft}Pre\ {\isacharprime}{\isacharprime}A{\isacharprime}{\isacharprime}\ {\isacharbrackleft}Var\ {\isadigit{1}}{\isacharcomma}\ Var\ {\isadigit{0}}{\isacharbrackright}{\isacharparenright}{\isacharparenright}{\isacharparenright}{\isacartoucheclose}\isanewline
\isadelimproof
\ \ %
\endisadelimproof
\isatagproof
\isacommand{using}\isamarkupfalse%
\ as{\isacharunderscore}given{\isacharunderscore}by{\isacharunderscore}the{\isacharunderscore}user{\isacharunderscore}any{\isacharunderscore}unis\ put{\isacharunderscore}unis{\isachardot}simps\ \isacommand{by}\isamarkupfalse%
\ metis%
\endisatagproof
{\isafoldproof}%
\isadelimproof
\endisadelimproof
}
\DefineSnippet{ex2cor2}{
\isacommand{corollary}\isamarkupfalse%
\ without{\isacharunderscore}two{\isacharunderscore}Uni{\isacharcolon}\isanewline
\ \ {\isacartoucheopen}semantics\ e\ f\ g\ {\isacharparenleft}Imp\ {\isacharparenleft}Uni\ {\isacharparenleft}Uni\ {\isacharparenleft}Pre\ {\isacharprime}{\isacharprime}A{\isacharprime}{\isacharprime}\ {\isacharbrackleft}Var\ {\isadigit{1}}{\isacharcomma}\ Var\ {\isadigit{0}}{\isacharbrackright}{\isacharparenright}{\isacharparenright}{\isacharparenright}\ {\isacharparenleft}Pre\ {\isacharprime}{\isacharprime}A{\isacharprime}{\isacharprime}\ {\isacharbrackleft}Var\ {\isadigit{1}}{\isacharcomma}\ Var\ {\isadigit{0}}{\isacharbrackright}{\isacharparenright}{\isacharparenright}{\isacartoucheclose}\isanewline
\isadelimproof
\ \ %
\endisadelimproof
\isatagproof
\isacommand{using}\isamarkupfalse%
\ as{\isacharunderscore}given{\isacharunderscore}by{\isacharunderscore}the{\isacharunderscore}user{\isacharunderscore}any{\isacharunderscore}unis\ put{\isacharunderscore}unis{\isachardot}simps\ \isacommand{by}\isamarkupfalse%
\ metis%
\endisatagproof
{\isafoldproof}%
\isadelimproof
\endisadelimproof
}
\DefineSnippet{ex2end}{
\isadelimtheory
\endisadelimtheory
\isatagtheory
\isacommand{end}\isamarkupfalse%
\endisatagtheory
{\isafoldtheory}%
\isadelimtheory
\endisadelimtheory
}

%% file: main.bbl
\begin{thebibliography}{10}
\providecommand{\bibitemdeclare}[2]{}
\providecommand{\surnamestart}{}
\providecommand{\surnameend}{}
\providecommand{\urlprefix}{Available at }
\providecommand{\url}[1]{\texttt{#1}}
\providecommand{\href}[2]{\texttt{#2}}
\providecommand{\urlalt}[2]{\href{#1}{#2}}
\providecommand{\doi}[1]{doi:\urlalt{http://dx.doi.org/#1}{#1}}
\providecommand{\bibinfo}[2]{#2}

\bibitemdeclare{misc}{proofweb}
\bibitem{proofweb}
\emph{\bibinfo{title}{ProofWeb. (ProofWeb is both a system for teaching logic
  and for using proof assistants through the web).}}
\newblock \urlprefix\url{http://proofweb.cs.ru.nl/login.php}.
\newblock \bibinfo{note}{Accessed December 2017}.

\bibitemdeclare{article}{Berghofer}
\bibitem{Berghofer}
\bibinfo{author}{Stefan \surnamestart Berghofer\surnameend}
  (\bibinfo{year}{2007}): \emph{\bibinfo{title}{First-Order Logic According to
  Fitting}}.
\newblock {\sl \bibinfo{journal}{Archive of Formal Proofs}}.
\newblock \bibinfo{note}{\url{http://isa-afp.org/entries/FOL-Fitting.html},
  Formal proof development}.

\bibitemdeclare{inproceedings}{Blanchette}
\bibitem{Blanchette}
\bibinfo{author}{Jasmin~Christian \surnamestart Blanchette\surnameend},
  \bibinfo{author}{Andrei \surnamestart Popescu\surnameend} \&
  \bibinfo{author}{Dmitriy \surnamestart Traytel\surnameend}
  (\bibinfo{year}{2014}): \emph{\bibinfo{title}{Unified Classical Logic
  Completeness - {A} Coinductive Pearl}}.
\newblock {\sl \bibinfo{series}{Lecture Notes in Computer Science}}
  \bibinfo{volume}{8562}, \bibinfo{publisher}{Springer}, pp.
  \bibinfo{pages}{46--60}, \doi{10.1007/978-3-319-08587-6_4}.

\bibitemdeclare{inproceedings}{breitner}
\bibitem{breitner}
\bibinfo{author}{Joachim \surnamestart Breitner\surnameend}
  (\bibinfo{year}{2016}): \emph{\bibinfo{title}{Visual Theorem Proving with the
  Incredible Proof Machine}}.
\newblock In: {\sl \bibinfo{booktitle}{Interactive Theorem Proving - 7th
  International Conference, {ITP} 2016, Nancy, France, August 22-25, 2016,
  Proceedings}}, pp. \bibinfo{pages}{123--139},
  \doi{10.1007/978-3-319-43144-4_8}.

\bibitemdeclare{article}{IncredibleProofMachineAFP}
\bibitem{IncredibleProofMachineAFP}
\bibinfo{author}{Joachim \surnamestart Breitner\surnameend} \&
  \bibinfo{author}{Denis \surnamestart Lohner\surnameend}
  (\bibinfo{year}{2016}): \emph{\bibinfo{title}{The meta theory of the
  Incredible Proof Machine}}.
\newblock {\sl \bibinfo{journal}{Archive of Formal Proofs}}.
\newblock
  \bibinfo{note}{\url{http://isa-afp.org/entries/Incredible_Proof_Machine.html},
  Formal proof development}.

\bibitemdeclare{article}{pandora}
\bibitem{pandora}
\bibinfo{author}{Krysia \surnamestart Broda\surnameend},
  \bibinfo{author}{Jiefei \surnamestart Ma\surnameend},
  \bibinfo{author}{Gabrielle \surnamestart Sinnadurai\surnameend} \&
  \bibinfo{author}{Alexander~J. \surnamestart Summers\surnameend}
  (\bibinfo{year}{2007}): \emph{\bibinfo{title}{Pandora: {A} Reasoning Toolbox
  using Natural Deduction Style}}.
\newblock {\sl \bibinfo{journal}{Logic Journal of the {IGPL}}}
  \bibinfo{volume}{15}(\bibinfo{number}{4}), pp. \bibinfo{pages}{293--304},
  \doi{10.1093/jigpal/jzm020}.

\bibitemdeclare{inproceedings}{Elsman}
\bibitem{Elsman}
\bibinfo{author}{Martin \surnamestart Elsman\surnameend}
  (\bibinfo{year}{2011}): \emph{\bibinfo{title}{SMLtoJs: Hosting a Standard ML
  Compiler in a Web Browser}}.
\newblock In: {\sl \bibinfo{booktitle}{Proceedings of the 1st ACM SIGPLAN
  International Workshop on Programming Language and Systems Technologies for
  Internet Clients}}, \bibinfo{series}{PLASTIC '11}, \bibinfo{publisher}{ACM},
  \bibinfo{address}{New York, NY, USA}, pp. \bibinfo{pages}{39--48},
  \doi{10.1145/2093328.2093336}.

\bibitemdeclare{book}{Fitting}
\bibitem{Fitting}
\bibinfo{author}{Melvin \surnamestart Fitting\surnameend}
  (\bibinfo{year}{1996}): \emph{\bibinfo{title}{First-Order Logic and Automated
  Theorem Proving, Second Edition}}.
\newblock \bibinfo{series}{Graduate Texts in Computer Science},
  \bibinfo{publisher}{Springer}, \doi{10.1007/978-1-4612-2360-3}.

\bibitemdeclare{}{From}
\bibitem{From}
\bibinfo{author}{Andreas~Halkj{\ae}r \surnamestart From\surnameend}
  (\bibinfo{year}{2017}): \emph{\bibinfo{title}{Formalized First-Order Logic}}.
\newblock \bibinfo{note}{BSc Thesis, Technical University of Denmark}.

\bibitemdeclare{inproceedings}{panda}
\bibitem{panda}
\bibinfo{author}{Olivier \surnamestart Gasquet\surnameend},
  \bibinfo{author}{Fran{\c{c}}ois \surnamestart Schwarzentruber\surnameend} \&
  \bibinfo{author}{Martin \surnamestart Strecker\surnameend}
  (\bibinfo{year}{2011}): \emph{\bibinfo{title}{Panda: {A} Proof Assistant in
  Natural Deduction for All. {A} Gentzen Style Proof Assistant for
  Undergraduate Students}}.
\newblock {\sl \bibinfo{series}{Lecture Notes in Computer Science}}
  \bibinfo{volume}{6680}, \bibinfo{publisher}{Springer}, pp.
  \bibinfo{pages}{85--92}, \doi{10.1007/978-3-642-21350-2_11}.

\bibitemdeclare{inproceedings}{Harrison}
\bibitem{Harrison}
\bibinfo{author}{John \surnamestart Harrison\surnameend}
  (\bibinfo{year}{1998}): \emph{\bibinfo{title}{Formalizing Basic First Order
  Model Theory}}.
\newblock In: {\sl \bibinfo{booktitle}{Theorem Proving in Higher Order Logics,
  11th International Conference, TPHOLs'98, Canberra, Australia, September 27 -
  October 1, 1998, Proceedings}}, \bibinfo{publisher}{Springer}, pp.
  \bibinfo{pages}{153--170}, \doi{10.1007/BFb0055135}.

\bibitemdeclare{book}{Huth}
\bibitem{Huth}
\bibinfo{author}{Michael \surnamestart Huth\surnameend} \&
  \bibinfo{author}{Mark \surnamestart Ryan\surnameend} (\bibinfo{year}{2004}):
  \emph{\bibinfo{title}{Logic in Computer Science: Modelling and Reasoning
  about Systems. Second Edition}}.
\newblock \bibinfo{publisher}{Cambridge University Press},
  \doi{10.1017/CBO9780511810275}.

\bibitemdeclare{article}{FOLHarrisonAFP}
\bibitem{FOLHarrisonAFP}
\bibinfo{author}{Alexander~Birch \surnamestart Jensen\surnameend},
  \bibinfo{author}{Anders \surnamestart Schlichtkrull\surnameend} \&
  \bibinfo{author}{J{\o}rgen \surnamestart Villadsen\surnameend}
  (\bibinfo{year}{2017}): \emph{\bibinfo{title}{First-Order Logic According to
  Harrison}}.
\newblock {\sl \bibinfo{journal}{Archive of Formal Proofs}}.
\newblock \bibinfo{note}{\url{http://isa-afp.org/entries/FOL_Harrison.html},
  Formal proof development}.

\bibitemdeclare{inproceedings}{Kumar}
\bibitem{Kumar}
\bibinfo{author}{Ramana \surnamestart Kumar\surnameend}, \bibinfo{author}{Rob
  \surnamestart Arthan\surnameend}, \bibinfo{author}{Magnus~O. \surnamestart
  Myreen\surnameend} \& \bibinfo{author}{Scott \surnamestart Owens\surnameend}
  (\bibinfo{year}{2014}): \emph{\bibinfo{title}{{HOL} with Definitions:
  Semantics, Soundness, and a Verified Implementation}}.
\newblock {\sl \bibinfo{series}{Lecture Notes in Computer Science}}
  \bibinfo{volume}{8858}, \bibinfo{publisher}{Springer}, pp.
  \bibinfo{pages}{308--324}, \doi{10.1007/978-3-319-08970-6_20}.

\bibitemdeclare{book}{Klein}
\bibitem{Klein}
\bibinfo{author}{Tobias \surnamestart Nipkow\surnameend} \&
  \bibinfo{author}{Gerwin \surnamestart Klein\surnameend}
  (\bibinfo{year}{2014}): \emph{\bibinfo{title}{Concrete Semantics - With
  Isabelle/HOL}}.
\newblock \bibinfo{publisher}{Springer}, \doi{10.1007/978-3-319-10542-0}.

\bibitemdeclare{book}{Nipkow}
\bibitem{Nipkow}
\bibinfo{author}{Tobias \surnamestart Nipkow\surnameend},
  \bibinfo{author}{Lawrence~C. \surnamestart Paulson\surnameend} \&
  \bibinfo{author}{Markus \surnamestart Wenzel\surnameend}
  (\bibinfo{year}{2002}): \emph{\bibinfo{title}{Isabelle/HOL - {A} Proof
  Assistant for Higher-Order Logic}}.
\newblock {\sl \bibinfo{series}{Lecture Notes in Computer Science}}
  \bibinfo{volume}{2283}, \bibinfo{publisher}{Springer},
  \doi{10.1007/3-540-45949-9}.

\bibitemdeclare{inproceedings}{kjs}
\bibitem{kjs}
\bibinfo{author}{Daejun \surnamestart Park\surnameend}, \bibinfo{author}{Andrei
  \surnamestart Stef\u{a}nescu\surnameend} \& \bibinfo{author}{Grigore
  \surnamestart Ro\c{s}u\surnameend} (\bibinfo{year}{2015}):
  \emph{\bibinfo{title}{KJS: A Complete Formal Semantics of JavaScript}}.
\newblock In: {\sl \bibinfo{booktitle}{Proceedings of the 36th ACM SIGPLAN
  Conference on Programming Language Design and Implementation}},
  \bibinfo{series}{PLDI '15}, \bibinfo{publisher}{ACM}, \bibinfo{address}{New
  York, NY, USA}, pp. \bibinfo{pages}{346--356}, \doi{10.1145/2737924.2737991}.

\bibitemdeclare{article}{Seldin}
\bibitem{Seldin}
\bibinfo{author}{Jonathan~P. \surnamestart Seldin\surnameend}
  (\bibinfo{year}{1989}): \emph{\bibinfo{title}{Normalization and excluded
  middle. {I}}}.
\newblock {\sl \bibinfo{journal}{Studia Logica}}
  \bibinfo{volume}{48}(\bibinfo{number}{2}), pp. \bibinfo{pages}{193--217},
  \doi{10.1007/BF02770512}.

\bibitemdeclare{article}{IFCoLog}
\bibitem{IFCoLog}
\bibinfo{author}{J{\o}rgen \surnamestart Villadsen\surnameend},
  \bibinfo{author}{Alexander~Birch \surnamestart Jensen\surnameend} \&
  \bibinfo{author}{Anders \surnamestart Schlichtkrull\surnameend}
  (\bibinfo{year}{2017}): \emph{\bibinfo{title}{{NaDeA}: {A} Natural Deduction
  Assistant with a Formalization in Isabelle}}.
\newblock {\sl \bibinfo{journal}{IFCoLog Journal of Logics and their
  Applications}} \bibinfo{volume}{4}(\bibinfo{number}{1}), pp.
  \bibinfo{pages}{55--82}.

\bibitemdeclare{misc}{Isar}
\bibitem{Isar}
\bibinfo{author}{Makarius \surnamestart Wenzel\surnameend}
  (\bibinfo{year}{2017}): \emph{\bibinfo{title}{The {Isabelle/Isar} Reference
  Manual}}.
\newblock
  \bibinfo{note}{\url{http://isabelle.in.tum.de/dist/doc/isar-ref.pdf}}.

\end{thebibliography}
